\documentclass[twocolumn,apl,amsmath,amssymb,showpacs,superscriptaddress]{revtex4-1}
\usepackage{epsf}      
\usepackage{graphicx}
\usepackage{color}
\usepackage{soul}
\usepackage{gensymb}
\usepackage{sidecap}
\usepackage{amsmath}
\usepackage{mathtools}
\usepackage[hidelinks,colorlinks=true,linkcolor=blue,citecolor=blue]{hyperref}

\begin{document}

\title{Correlation between the exchange bias effect and antisite disorders in Sr$_{2-x}$La$_x$CoNbO$_6$}

\author{Ajay Kumar}
\affiliation{Department of Physics, Indian Institute of Technology Delhi, Hauz Khas, New Delhi-110016, India}
\author{B. Schwarz}
\affiliation{Institute for Applied Materials (IAM), Karlsruhe Institute of Technology (KIT), 76344, Eggenstein-Leopoldshafen, Germany}
\author{R. S. Dhaka}
\email{rsdhaka@physics.iitd.ac.in}
\affiliation{Department of Physics, Indian Institute of Technology Delhi, Hauz Khas, New Delhi-110016, India}

\date{\today}

\begin{abstract}

We unravel the effect of La substitution and hence antisite disorders on the exchange bias (EB) mechanism in Sr$_{2-x}$La$_x$CoNbO$_6$ ($x=$ 0, 0.2) double perovskite samples using the detailed analysis of the field cooled magnetization isotherms (M--H) and training effect. The field dependence of the freezing temperature deviates from both Gabay-Toulouse (GT) and de Almeida-Thouless (AT) lines and the analysis suggests that the $x=$ 0 sample follows a different universality class with moderate anisotropy in the frozen spins. Interestingly, we find that the EB effect is significantly suppressed in the $x=$ 0.2 sample due to increase [decrease] in the size of ferromagnetic (FM) [cluster glass (CG)] domain, which reduces the effective disordered interface responsible for the EB. The changes in fraction of FM, AFM, and CG like interactions with the La substitution and applied magnetic field are found to be crucial in governing the EB effect in these samples. Further, the training effect measurements show the unequal shift in the left and right branches of the M--H loops and their different evolution with the field cycles ($n$). The analysis reveals that the rotatable spins relax approximately one order of magnitude faster than the frozen spins at the disordered interface. We find a possible correlation between the observed EB effect and the antisite disorders in these samples. 

\end{abstract}

\maketitle
\section{\noindent ~Introduction}

The exchange bias (EB) effect arises from the unidirectional magnetic anisotropy resulting from the exchange interactions at the ferromagnetic (FM)-antiferromagnetic (AFM) interface and is manifested by a horizontal and/or vertical shift in the M--H loop, recorded after cooling the sample in the applied magnetic field below the AFM transition temperature (T$_{\rm N}$) \cite{Meiklejohn_PR_56, Stamps_JD_2000}. In addition to the FM--AFM interface, the EB effect has also been observed at the FM-ferrimagnetic (FiM), AFM--FiM, FM--spin glass (SG), AFM--SG, and FiM-SG disordered interfaces due to frustration in the spins \cite{Vasilakaki_PRB_09, Berkowitz_PRB_08, Karmakar_PRB_08, Sahoo_PRB_19, Anusree_PRB_20}. The materials showing the EB effect are being extensively studied due to their wide range of technological applications in spin valves \cite{Kuncser_JMMM_08}, hard magnets \cite{Sort_PRB_02}, magnetic random access memories (MRAM)  \cite{Ortega_PR_15, He_NM_10, Liu_AFM_11}, and various spintronic devices \cite{Wolf_Science_01, Binek_JAP_05, Murphy_ACSCS_21}. The degree and direction of the EB effect are governed by the nature of the exchange interactions at the interface. Thus, a precise understanding and control on the various exchange parameters at the magnetically disordered interfaces are necessary to engineer the suitable candidates for device applications. Also, the EB effect involving the spin glassy phase(s) exhibits the more complex phenomena due to the presence of the intrinsically frozen spins and is crucial for their fundamental and technological aspects \cite{Karmakar_PRB_08, Sahoo_PRB_19, Anusree_PRB_20, Ali_NM_07}. For example, the cluster glass (CG) behavior was recently found to govern the giant exchange bias of $\approx$9.6~kOe in SrLaFe$_{0.5}$Mn$_{0.25}$Co$_{0.25}$O$_4$ sample for a cooling field of 50~kOe  \cite{Anusree_PRB_20}. 

In this context, the Co--based perovskite oxides have gained huge attention due to various possible valence and spin states of Co in the octahedral environment, resulting in several competing magnetic interactions and hence peculiar magnetic ground states such as spin glass, cluster glass, spin liquid, spin ice, Griffiths phase etc. \cite{Motohashi_PRB_05, Fita_PRB_18, Shukla_PRB_18, Shukla_JPCC_19, Yan_PRL_14, Shukla_PRB_2023}. These low temperature frustrated magnetic states can be systematically manipulated by the external perturbations such as temperature, mechanical pressure, chemical pressure (doping), magnetic field, etc. \cite{Motohashi_PRB_05, Fita_PRB_18, Shukla_JPCC_19}. Further, the double perovskite (DP) oxides having general formula A$_2$BB$^\prime$O$_6$ (A is the rare earth/alkali earth metals, and B and B$^\prime$ are the transition metals) can crystallize either in the ordered configuration with the alternating BO$_6$ and B$^\prime$O$_6$ octahedra at the corners, or in the disordered state with two consecutive BO$_6$ or B$^\prime$O$_6$ octahedra. These configurations give rise to B-O-B$^\prime$-O-B and B$^\prime$-O-B-O-B$^\prime$, and B-O-B and B$^\prime$-O-B$^\prime$ exchange interactions, respectively \cite{Vasala_SSC_15}. Therefore, the fraction of various exchange interactions in DPs can be precisely controlled by varying the structural ordering, which is mainly governed by the ionic and valence mismatch between the two B-site cations \cite{Vasala_SSC_15, Kumar_PRB_19, Haripriya_PRB_19, Kumar_PRB1_20, Das_PRB_20, Bos_PRB_04}, as well as strain \cite{AjayJAP20, AjayJVSTA23}. For instance, the antisite (B-site) disorders in LaSrCoFeO$_6$ give rise to the Co$^{3+}$-Co$^{3+}$, Co$^{3+}$-Co$^{4+}$, Co$^{4+}$-Co$^{4+}$, Fe$^{3+}$-Fe$^{3+}$, Fe$^{3+}$-Fe$^{4+}$, and Fe$^{4+}$-Fe$^{4+}$ AFM exchange interactions, whereas the ordered domains exhibit Co$^{3+}$-Fe$^{4+}$ and Co$^{4+}$-Fe$^{3+}$ FM couplings \cite{ Sahoo_PRB_19}. These competing AFM-FM interactions result in the low-temperature SG behavior and a large EB effect in this sample \cite{ Sahoo_PRB_19}. The spin glass-like phase is reported to play a key role in governing the largest zero field cooled EB effect in La$_{1.5}$$A_{0.5}$CoMnO$_6$ ($A=$ Ba, Ca, Sr) samples \cite{Coutrim_PRB_18, Coutrim_PRB_19}. In addition, the antisite disorder induced glassy behavior and the EB effect were observed at the interface of the long-range FM/AFM ordered and glassy phases in Gd$_2$CoRuO$_6$ \cite{Das_PRB_20}, Sr$_2$FeCoO$_6$ \cite{Pradheesh_APL_12}, Pr$_2$CoMnO$_6$ \cite{Liu_JAP_14}, La$_{1.5}$Ca$_{0.5}$CoIrO$_6$ \cite{Coutrim_PRB_16}, La$_{2-x}$Sr$_x$CoMnO$_6$ \cite{Murthy_JPCM_16}, Eu$_2$CoMnO$_6$ \cite{Banerjee_PRB_18} etc. 

The Sr$_2$CoNbO$_6$ is one of the best prototype for understanding the complex EB mechanism in the magnetically disordered DPs due to the presence of only one magnetic cation (as Nb$^{5+}$; 4$d^0$ remains non-magnetic) and delicate antisite disorders in this sample \cite{Kumar_PRB1_20, Azcondo_DT_15}. The moderate ionic and valence mismatch between the Co$^{3+}$ and Nb$^{5+}$ ions \cite{Shannon_AC_76, Vasala_SSC_15} give rise to a site ordering with a correlation length of $<$100~\AA~\cite{Azcondo_DT_15}, but a macroscopically disordered crystal structure \cite{Kumar_PRB1_20, Kumar_PRB3_22, Azcondo_DT_15}, resulting in the intriguing magnetic ground state of Sr$_2$CoNbO$_6$ \cite{Kumar_PRB1_20, Kumar_PRB2_20, Azcondo_DT_15}. Also, the low temperature spin glass dynamics was reported in the Sr$_2$CoNbO$_6$ based on the memory effect and ac susceptibility data \cite{Azcondo_DT_15}. These investigations were further supported by our recent thermoremanent magnetization (TRM) and the aging effect measurements performed on Sr$_{2-x}$La$_x$CoNbO$_6$ samples \cite{Kumar_PRB2_20}. The detailed analysis of the  ac susceptibility data indicate the cluster glass like behavior in the Sr$_{2-x}$La$_x$CoNbO$_6$ samples for $x \leqslant$0.4, where the size of the glassy cluster, inter-cluster interactions, and spin-spin correlation strength within the cluster decreases with the La substitution \cite{Kumar_PRB2_20}. Moreover, the x-ray diffraction (XRD), neutron diffraction (ND), and extended x-ray absorption fine structure (EXAFS) measurements indicates a monotonic enhancement in the B-site ordering with La substitution in Sr$_{2-x}$La$_x$CoNbO$_6$, which is found to play an important role in governing the low temperature complex spin dynamics in these samples \cite{Kumar_PRB1_20, Kumar_PRB3_22, Kumar_JPCL_22}. Therefore, it is vital to explore the possibility of the EB effect resulting from several competing magnetic interactions in these samples and their direct correlation with the degree of B-site ordering. Such a correlation can lead to a much deeper insight of the complex EB mechanism in the DPs and will be useful in engineering the suitable candidates for the desired applications. 

In this paper, we present the effect of antisite disorders on the frozen spins and hence EB phenomena in Sr$_{2-x}$La$_x$CoNbO$_6$ ($x=$ 0, 0.2) samples. In the first section, we investigate the field dependence of the irreversibility temperature (T$_{\rm irr}$) in the zero-field-cooled (ZFC) and field-cooled (FC) magnetic susceptibility curves for the $x=$ 0 sample, which shows deviation from both Gabay-Toulouse (GT) and de Almeida–Thouless (AT) lines, and the analysis suggests the moderate exchange anisotropy in this sample. In the next section, we present the field dependence of the EB effect in the $x=$ 0 and 0.2 samples and found that the EB parameters increase with the cooling field (H$_{\rm CF}$) up to a certain value and then decreases with further increases in the H$_{\rm CF}$ for both the samples. More importantly, the La substitution significantly reduces  the EB effect due to enhancement in the size of the FM domains, which reduces the effective disordered interface. Then, we performed the training effect measurements on both the samples and use various models to understand the low temperature complex spin dynamics. We observe the asymmetric training effect in both the samples and a more stable EB mechanism in case of the $x=$ 0.2 sample. In the last section, we discuss the possible origins and evolution of the EB effect and their correlation with the antisite disorders in these samples. 

\section{\noindent ~Experimental}

The polycrystalline samples of Sr$_{2-x}$La$_{x}$CoNbO$_{6}$ ($x=$ 0 and 0.2) were synthesized by usual solid-state reaction route. The detailed synthesis process and physical properties can be found in our recent papers \cite{Kumar_PRB1_20, Kumar_PRB2_20, Kumar_PRB3_22, Kumar_JPCL_22}. The magnetization measurements were performed using the Physical Property Measurement System (PPMS) from Quantum Design, USA. The details of the protocols for different measurements are provided along with their respective discussion in the next section. 

\section{\noindent ~Results and discussion}

\subsection{\noindent ~Field dependence of the blocking temperature}

\begin{figure}
\includegraphics[width=0.5\textwidth]{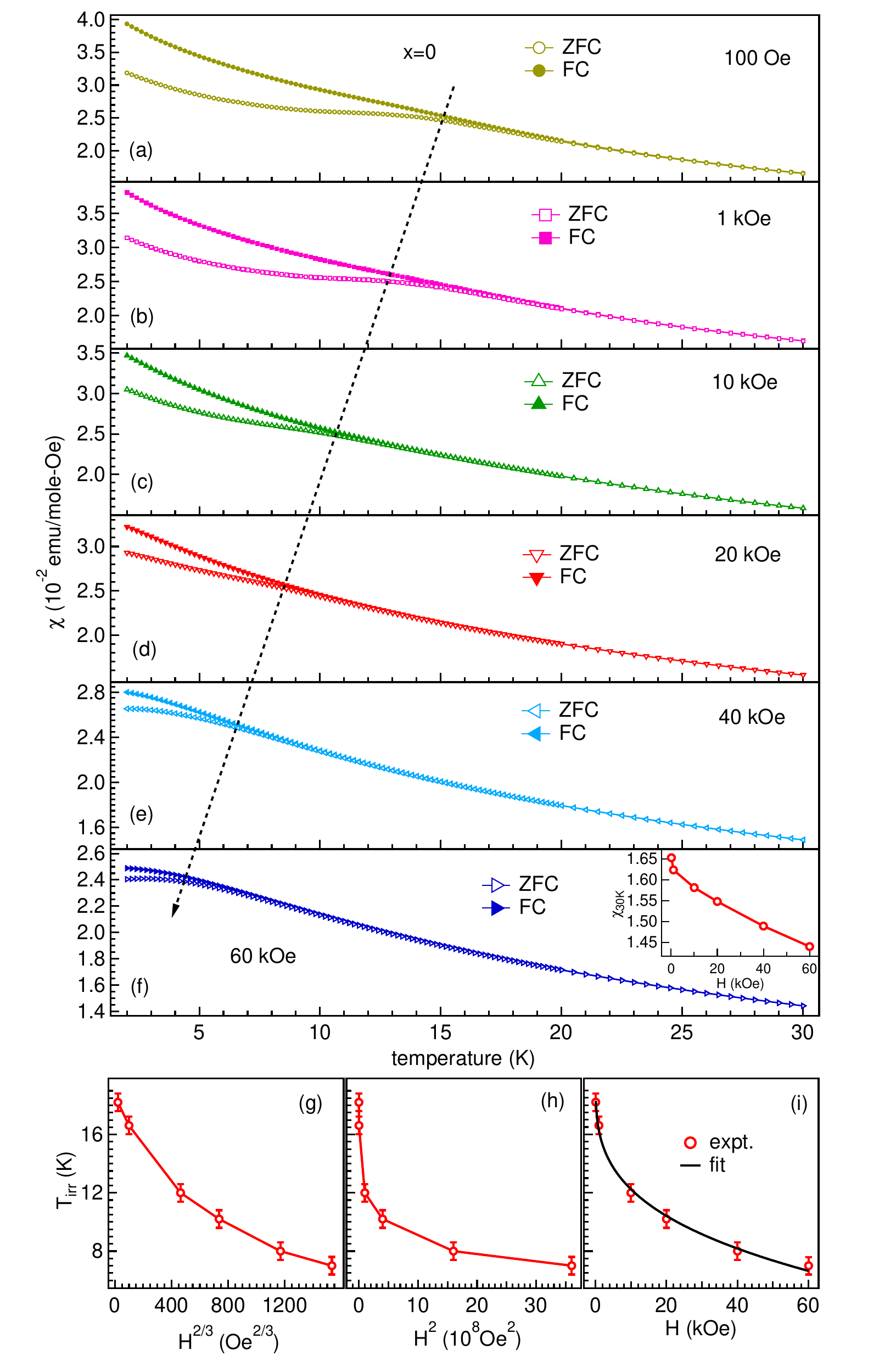}
\caption {(a--f) The ZFC and FC magnetic susceptibility curves recorded at different magnetic fields ($H$) for the $x=$ 0 sample, where an arrow represents the shift in the irreversibility temperature (T$_{\rm irr}$). Inset in (f) shows the field dependence of the magnetic susceptibility at 30~K. The variation of T$_{\rm irr}$ with (g) H$^{2/3}$, (h) H$^{2}$, and (i) H (see text for more details). The black solid line in (i) represents the best fit using equation 1.}
 \label{A1}
\end{figure}

In Figs.~\ref{A1}(a--f), we show the zero field cooled (ZFC) and field cooled (FC) magnetic susceptibility ($\chi$=M/H) curves of the $x=$0 sample, recorded at different magnetic fields to probe the low temperature complex spin dynamics \cite{Kumar_PRB1_20, Kumar_PRB2_20}. A clear bifurcation in the two curves is observed below 15~K at 100~Oe, which shifts to the lower temperature with increase in the applied magnetic field, as indicated by the dashed arrow in Fig.~\ref{A1}. This is the typical signature of the magnetically disordered systems with glassy ground state, where the higher applied magnetic field align even the blocked spins at the low temperature \cite{Kumar_PRB_19, Binder_RMP_86, Gabay_PRL_81, Cragg_PRL_82, Cragg_PRL2_82}. Here, it is interesting to note that the ZFC curves show a decreasing tendency below a certain temperature (T$_{\rm irr}$) and then increases with further decrease in the temperature for the low applied fields [see Figs.~\ref{A1} (a--d) up to 20~kOe], unlike conventional glassy systems, where these curves decrease monotonically on lowering the temperature \cite{Bag_PRB_18, Kumar_PRB_19}. However, this upturn in the ZFC curves below 5~K gradually becomes downturn at the higher applied magnetic fields $\ge$20~kOe [see Figs.~\ref{A1}(e, f)], which rules out the paramagnetic impurities as the origin of this low temperature upturn for H$\leqslant$ 20~kOe. The similar temperature and field dependence of the ZFC behavior is also reported in the several other disordered compounds \cite{Shand_PRB_10, Wakimoto_PRB_2000, Senchuk_EPJB_04}. In the present case, an enhancement in the fraction of FM component and hence weakly saturating behavior of the M--H curves (presented in next subsection) results in this low temperature downturn in the ZFC curves at the higher magnetic fields. This effect is more clearly observed in the exchange bias effect measurements performed at the higher cooling fields, discussed in detail below. Moreover, the magnetic susceptibility ($\chi$) monotonically decreases with  increase in the field even up to 30~K [see inset of Fig.~\ref{A1}(f) for the field dependence of $\chi$ at 30~K], which indicates the presence of short-range magnetic correlation in the sample well above the blocking temperature.  

Further, we analyze the magnetic field dependence of the irreversibility temperature in the ZFC--FC curves (T$_{\rm irr}$) using the following power law \cite{Malozemoff_PRL_83, Kaul_JPCM_98, Shand_PRB_10} 
\begin{equation}
T_{\rm irr}(H)= T_{\rm irr}(0)[1-{\rm A}H^{p}],
\label{Tirr}
\end{equation}
where T$_{\rm irr}$(0) is the irreversibility temperature in the zero field regime, A is a constant which is the measure of the anisotropy in the system, and the value of power exponent $p$ defines the nature of the magnetic interactions. For the highly anisotropic Ising spins, the value of $p$ is 2/3, which gives de Almeida-Thouless (A-T) line, whereas $p=$ 2 in the weak anisotropy regime corresponds to the Gabay-Toulouse (G-T) line \cite{Almeida_JPAMG_78, Gabay_PRL_81}. To check this, we plot the T$_{\rm irr}$ as a function of H$^{2/3}$ and H$^2$ in Figs.~\ref{A1}(g) and (h), respectively. Interestingly, we observe a clear deviation from the linearity in both the cases, which indicates that none of the two irreversibility lines reproduces the observed behavior in the present case. Therefore, we fit the T$_{\rm irr}$(H) curve using the equation~\ref{Tirr} by keeping the exponent $p$ as a free parameter, which gives A = 0.028$\pm$0.001 Oe$^{-0.29}$, $p=$ 0.29$\pm$0.03, and T$_{\rm irr}$(0) = 20.5$\pm$0.3~K and the best fit curve is shown by the black solid line in Fig.~\ref{A1}(i). The value of power exponent $p$ is significantly lower than both A-T and G-T lines, which indicate the existence of a different universality class in the present case. Kotliar and Sompolinsky theoretically redefine the A-T and G-T lines with the $D/k_BT>>(\mu H/k_BT)^{2/3}$ and $D/k_BT<<(\mu H/k_BT)^{5/2}$ conditions in the strong and week anisotropic regimes, respectively, where D is the strength of the Dzyaloshinskii-Moriya interactions, k$_B$ is the Boltzmann constant, and $\mu$ is the magnetic moment of a spin \cite{Kotliar_PRL_84}. Interestingly, the authors observed the value of power exponent $p=$ 1/3 in the moderate anisotropic regime, where $(\mu H/k_BT)^{5/2}<<D/k_BT<<\mu H/k_BT$ \cite{Kotliar_PRL_84}, which is  close to the value obtained in the present case. Thus, we believe that the $x=$ 0 sample follows a different universality class with an intermediate random anisotropy, lying between A-T and G-T lines. A similar behavior is also reported in the other disordered compounds such as Mn$_{0.25}$Ti$_{1.1}$S$_2$ \cite{Shand_PRB_10}, ZnTiCoO$_4$ \cite{Chowdhury_JPCM_22}, and Ba$_2$CoRuO$_6$ \cite{Chakravarty_PRB_23}. 

\begin{figure*}
\includegraphics[width=7.3in]{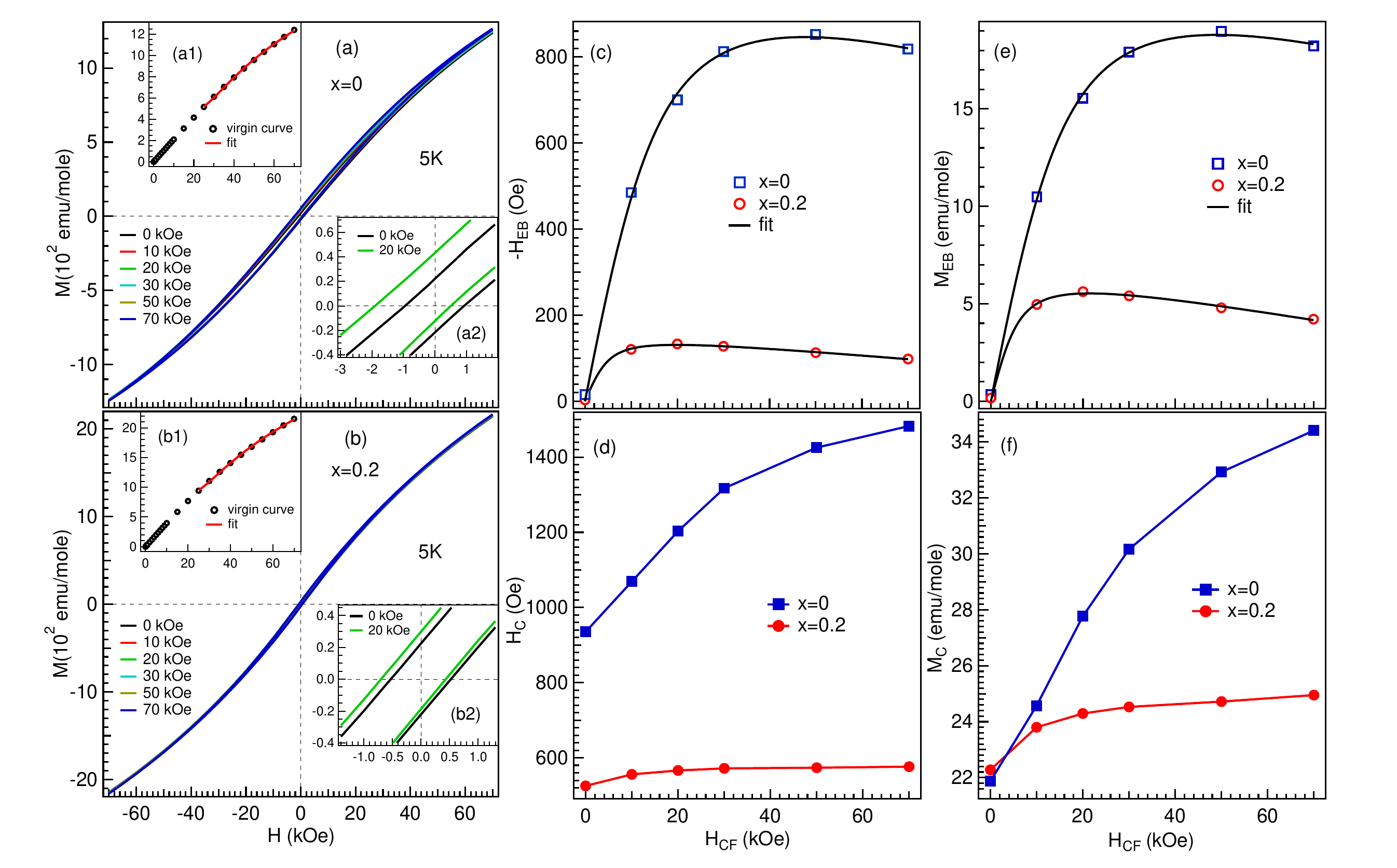}
\caption {The M--H curves recorded at 5~K in ZFC and FC modes for the (a) $x=$ 0 and (b) 0.2 samples. Insets (a1) and (b1) show the virgin magnetization isotherms in the ZFC mode with the best fit (solid red line) using equation~\ref{virgin} in the high field ($>$25~kOe) regime, and insets (a2) and (b2) are the enlarged view of the low field region for the zero field and 20~kOe field cooled modes, for the $x=$ 0 sand 0.2 samples, respectively. The cooling field (H$_{\rm CF}$) dependence of the (c) exchange bias field (H$_{\rm EB}$), (d) coercive field (H$_{\rm C}$), (e) exchange bias magnetization (M$_{\rm EB}$), and (f) magnetic coercivity (M$_{\rm C}$) at 5~K for the $x=$ 0 and 0.2 samples. The solid black lines in (c) and (e) represent the best fitted curves using equation~\ref{HEB2}. }
 \label{EB1}
\end{figure*}

\subsection{\noindent ~Isothermal magnetization and the EB effect}

In order to further unravel the low temperature spin dynamics, we measure the field dependent magnetization (M--H) curves between $\pm$70~kOe at 5~K for the $x=$ 0 and 0.2 samples, after cooling the samples in the different magnetic fields (H$_{\rm CF}$), as shown in Figs.~\ref{EB1}(a) and (b), respectively. The ZFC M--H curves show the coercivity of around 930~Oe and 520~Oe for the $x=$ 0 and 0.2 samples, indicating the presence of weak, but finite FM interactions in these samples, which are found to decrease with the La substitution. Further, the non-saturating behavior of the M--H curves even up to 70~kOe suggests the presence of canted spins and/or local AFM coupling in both the samples. To further understand this, we fit the ZFC virgin magnetization isotherms of both the samples in the high field region (H $ \geqslant 25$~kOe) using the law of approach to the saturation \cite{Andreev_Jalcom_97, Sahoo_PRB_19}
\begin{equation}
M (H)=M_S \left(1-\frac{a}{H}-\frac{b}{H^2}\right)+\chi_{\rm hf}H,
\label{virgin}
\end{equation}
where M$_{\rm S}$ is the saturation magnetization, terms $a/H$ and b/H$^2$ are attributed to the structural defects and magnetocrystalline anisotropy, respectively, and $\chi_{\rm hf}$ represents the high field susceptibility due to the enhanced spontaneous magnetization as a result of the applied magnetic field. The best fit curves for both the samples are shown by the solid red lines in the respective insets (a1) and (b1) of Figs.~\ref{EB1}(a, b), which gives M$_{\rm S}$=0.26 and 0.44~$\mu_B$/f.u. for the $x=$ 0 and 0.2 samples, respectively. The calculated values of the saturation magnetization are significantly lower than the theoretical saturation moment (M$_{\rm S}$) for free Co$^{3+}$/Co$^{2+}$ ions (g$_J$J$\mu_B$ = 6 $\mu_B$), which further confirm the local AFM coupling/canted spins and/or the strong crystal field in these samples. The higher value of M$_{\rm S}$ in case of the $x=$ 0.2 as compared to the $x=$ 0 sample suggests the stronger FM interactions in the former. Moreover, the Co$^{2+}$ in the octahedral environment shows significant unquenched orbital magnetic moment \cite{Mabbs_Dover_08, Lloret_ICA_08, Viola_CM_03}, which can also lead to this higher magnetic moment in case of the $x=$ 0.2 sample as La substitution converts 20\% Co from 3+ to 2+ \cite{Kumar_PRB1_20, Kumar_PRB3_22}. 

Here, it is important to emphasize again that the mixed FM--AFM interactions result in the low temperature cluster glass (CG) behavior of these samples \cite{Kumar_PRB2_20}. Further, the frustrated spins due to multiple magnetic interactions can lead to the exchange bias effect in the phase separated polycrystalline samples \cite{Sahoo_PRB_19, Anusree_PRB_20, Banerjee_PRM_23, Karmakar_PRB_08, Kumar_PRB_21}. In the present case, we observe the symmetric M--H loops around H=0 axis in the ZFC mode for both the samples, which indicate the absence of the anisotropy due to the randomness at the FM/AFM--CG boundaries. However, a significant shift of the M--H curves towards the negative field and positive magnetization axes is clearly observed in the FC modes, as shown in the insets (a2) and (b2) of the Figs.~\ref{EB1} (a, b) for the $x=$ 0 and 0.2 samples, respectively. This indicates the presence of the conventional EB effect in both the samples due to the unidirectional anisotropic exchange interactions \cite{Nogues_PR_05, Zhang_PRB_18, Anusree_PRB_20}. However, the $x=$ 0.2 sample shows much lower shift in the hysteresis loop as compared to the $x=$ 0 sample [see insets (a2) and (b2) of the Figs.~\ref{EB1}(a, b)], indicating the lower interfacial exchange coupling in this sample. A reduction in the size of the interacting spin cluster in case the $x=$ 0.2 as compared to the $x=$ 0 sample \cite {Kumar_PRB2_20} may be the possible reason for this observed reduction in the EB with the La substitution. 

Now we focus on the quantitative understanding of this observed shift in the hysteresis, both horizontally as well as vertically by calculating the exchange bias field [H$_{\rm EB}$=(H$_{\rm C+}$+H$_{\rm C-}$)/2], coercive field [H$_C$= ($|H_{\rm C+}|$+$|H_{\rm C-}|$)/2], exchange bias magnetization [M$_{\rm EB}$=(M$_{\rm C+}$+M$_{\rm C-}$)/2], and magnetic coercivity [M$_{\rm C}$= ($|M_{\rm C+}|$+$|M_{\rm C-}|$)/2], where H$_{\rm C+}$ and H$_{\rm C-}$ are the positive (right) and negative (left) coercive fields, and M$_{\rm C+}$ and M$_{\rm C-}$ are the positive (upper) and negative (lower) remanent magnetization, respectively \cite{Sahoo_PRB_19, Das_PRB_18}, as shown in Figs.~\ref{EB1}(c--f) for both the samples. Interestingly, all the four parameters decrease with the La substitution. More strikingly, the exchange bias field as well as exchange bias magnetization in case of the $x =$ 0 sample increases with increase in the cooling field, then attain a saturation value around 50~kOe, and then decreases with further increase in the H$_{\rm CF}$ for 70~kOe. On the other hand, both of these parameters attain a saturation value at a much lower H$_{\rm CF}$ of around 20~kOe in case of the $x=$ 0.2 sample, and then decreases for the higher values of H$_{\rm CF}$ [see Figs.~\ref{EB1}(c, e)]. A similar reduction in the exchange bias parameters at the higher cooling fields is also reported in the several phase separated polycrystalline samples, which is attributed to the growth of the FM domains \cite{Radu_PRB_03, Tang_PRB_06, Karmakar_PRB_08, Sahoo_PRB_19, Kumar_PRB_21, Banerjee_PRM_23}. The applied H$_{\rm CF}$ reduces the random direction of the anisotropic exchange interactions at the FM--AFM interface by aligning the FM spins in the direction of field. Thus, the H$_{\rm EB}$ and M$_{\rm EB}$ increases with increase in H$_{\rm CF}$. However, at the higher H$_{\rm CF}$ the FM/CG interactions dominate over the AFM coupling causing a reduction in the EB parameters. An increment in the FM interactions and/or growth of the FM domains at the higher field is evident from the enhancement in the H$_{\rm C}$ and M$_{\rm C}$ values up to 70~kOe for both the samples [see Figs.~\ref{EB1}(d, f)]. This can be quantitatively understood from the following EB relation developed for the AFM--FM film interfaces \cite{Meiklejohn_PR_56, Karmakar_PRB_08}
\begin{equation}
-H_{\rm EB}=J \frac{S_{\rm AFM}S_{\rm FM}}{\mu_0t_{\rm FM}M_{\rm FM}}, 
\label{EB}
\end{equation}
where J is the exchange integral per unit area at the FM--AFM interface, S$_{\rm AFM}$ and S$_{\rm FM}$ are the magnetization at the AFM and FM interfaces, respectively, $t_{\rm FM}$ and M$_{\rm FM}$ are the thickness and magnetization of the FM film, respectively. The increase in H$_{\rm CF}$ enhances the interfacial magnetization S$_{\rm AFM}$ and S$_{\rm FM}$ as well as $t_{\rm FM}$ and M$_{\rm FM}$ due to growth and alignment of the FM domains, respectively, which causes an overall enhancement in the H$_{\rm EB}$. However, at the higher H$_{\rm CF}$, the FM clusters grow in size, which decreases the FM--AFM interface area and hence the values of S$_{\rm AFM}$ and S$_{\rm FM}$ accompanied by enhancement in the $t_{\rm FM}$ and M$_{\rm FM}$. Therefore, the H$_{\rm EB}$ decreases for the strong enough H$_{\rm CF}$ (equation \ref{EB}). 

It is clear from the above discussion that the value of this critical cooling field (H$_{\rm CF}^{\rm crit}$) above which H$_{\rm EB}$ decreases depends on the strength of the exchange interactions at the FM--AFM interface. Therefore, to further understand this, the cooling field dependence of the EB field is expressed using the following relation, given by Niebieskikwiat and Salamon for $\mu_0$H$_{\rm CF} <$ k$_B$T \cite{Niebieskikwiat_PRB_05, Karmakar_PRB_08, Fita_PRB_18} 
\begin{equation}
-H_{EB} \propto J\left[\frac{J\mu_0}{\left(g\mu_B\right)^2}L \left(\frac{\mu H_{\rm CF}}{k_BT_f}\right)+H_{\rm CF}\right], 
\label{HEB2}
\end{equation}
where $J$ is the exchange integral at the FM--AFM interface, $\mu_0$=4.0 and 3.8$\mu_B$ for the $x=$ 0 and 0.2 samples, respectively [4.0~$\mu_B$ for HS Co$^{3+}$ (S=2) and 3.0~$\mu_B$ for HS Co$^{2+}$ (S=3/2) ions], $\mu_B$ is the Bohr magneton, $g=$ 2 is the Lande $g$-factor, $L$ represent the Langevin function, $\mu$=N$\mu_0$ is the ferromagnetic moment, N is the number of spins in a ferromagnetic cluster, $k_B$ is the Boltzmann's constant, and $T_f$ is the freezing temperature below which FM--AFM clusters coexist. For the smaller values of H$_{\rm CF}$ the first term in equation~\ref{HEB2} dominates, i.e., H$_{\rm EB}\propto$ J$^2$, whereas for the higher cooling field second term dominates, i.e., H$_{\rm EB}\propto$ J. 
\begin{figure}[h] 
\includegraphics[width=0.47\textwidth]{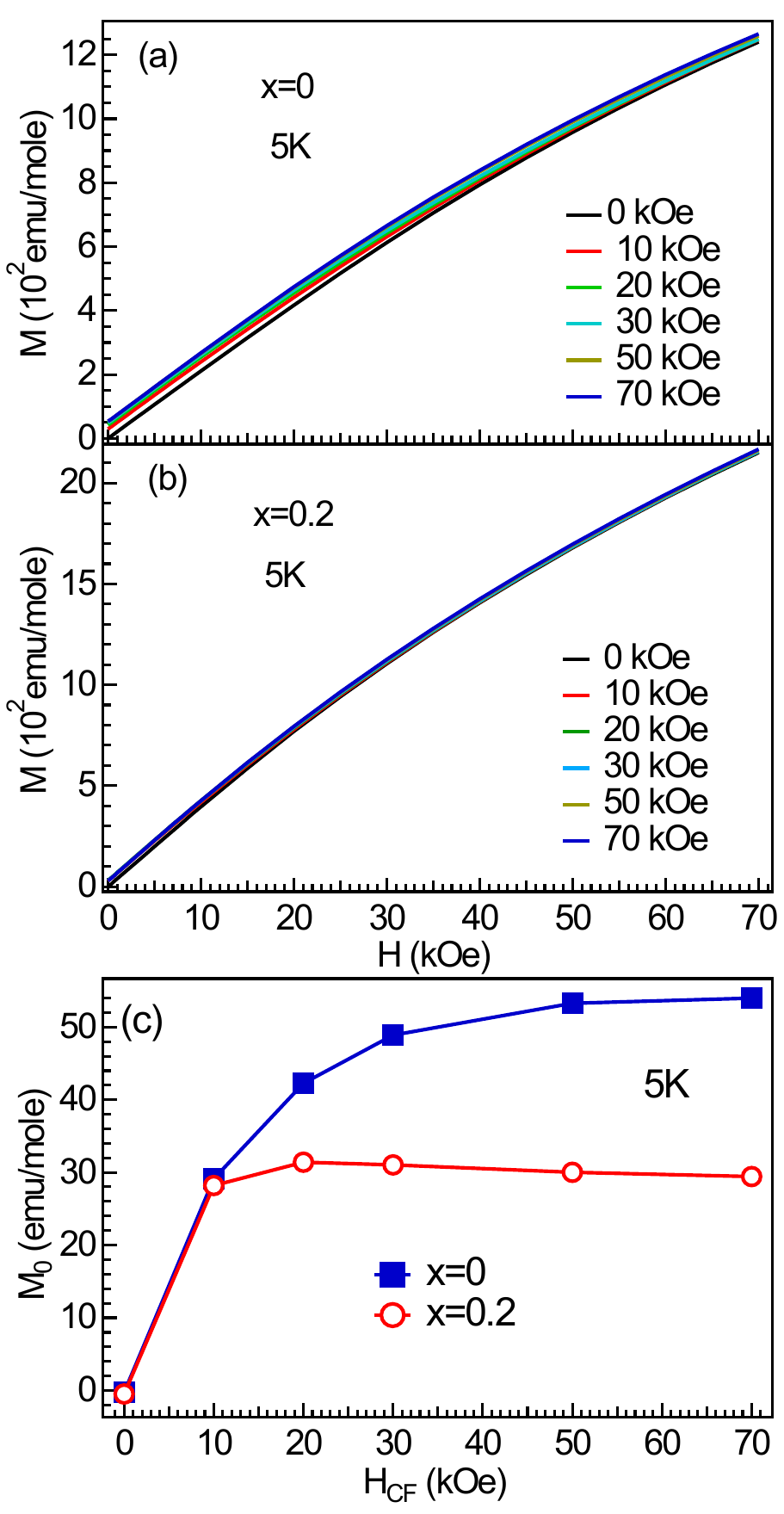}
\caption {(a, b) The virgin magnetization isotherms at 5~K in different field cooled states for the $x=$ 0 and 0.2 samples, respectively. (c) The zero field magnetization (M$_0$) as a function of the cooling field for both the samples.} 
\label{EB4_1}
\end{figure}
So, for the negative value of $J$, the H$_{\rm EB}$ is expected to first increase with the  H$_{\rm CF}$ and then decrease at the higher values where second term starts dominating over first, which can even change the sign of the H$_{\rm EB}$ \cite{Nogues_PRL_96, Nogues_PRB_2000, Sahoo_AMI_17, Hong_APL_12}. This indicates that the observed behavior of the  H$_{\rm EB}$ and M$_{\rm EB}$ in the present case can be well understood using the above model. Therefore, we fit the H$_{\rm EB}$ and M$_{\rm EB}$ as a function of H$_{\rm CF}$ using the  equation~\ref{HEB2} by taking T$_f$=13.5 and 9.7~K for the $x =$0 and 0.2, respectively \cite{Kumar_PRB2_20}, and varying N, $J$, and a proportionality constant. The best fit curves, represented by the solid black lines in Figs.~\ref{EB1}(c) and (e), exhibits an excellent agreement with the experimental data, which gives $J =$ -1.6$\pm$0.1 and -0.89$\pm$0.01~meV, and $N =$ 6$\pm$1 and 13$\pm$1 for the $x=$ 0 and 0.2 samples, respectively. It is important to note that the FM clusters are two times larger in case of the $x=$ 0.2 sample, which results in the smaller effective FM--AFM interface and hence a lower $J$ value as compared to the $x=$ 0 sample. These reduced FM--AFM boundaries which are necessary for the spin pinning and hence EB effect is responsible for the observed lower values of H$_{\rm EB}$ and M$_{\rm EB}$ in case of the $x=$ 0.2 sample. Moreover, for the $x=$ 0.2 sample the larger FM cluster results in the more prominent effect of the increase in H$_{\rm CF}$ and consequently reduction in H$_{\rm EB}$ and M$_{\rm EB}$ at a much lower H$_{\rm CF}$ as compared to the $x=$ 0. The negative $J$ value for both the samples indicate the presence of AFM coupling between the FM domains, causing the observed EB effect. 

The virgin magnetization isotherms are shown in Figs.~\ref{EB4_1}(a, b) for both the samples ($x=$ 0 and 0.2) after cooling them in the different magnetic fields. Note that cooling the samples in an applied magnetic field from 300~K (paramagnetic region, where spins are free to align) to 5~K results in the blocking of the aligned spins in the direction of H$_{\rm CF}$ due to their glassy ground state \cite{Kumar_PRB2_20}. This causes a remanent magnetization in these samples at H=0~Oe. This zero field magnetization (M$_0$) resulting from the blocked spins at the FM--AFM interface can be responsible for the observed EB effect in these samples. Fig.~\ref{EB4_1}(c) shows the variation of the zero field magnetization as a function of the cooling field, H$_{\rm CF}$. The M$_0$ of the $x=$ 0 sample increases rapidly with increase in the H$_{\rm CF}$ up to $\sim$30~kOe and then shows almost saturating behavior with further increase up to 70~kOe. On the other hand, the M$_0$ increases up to 20~kOe and then slightly but monotonically decreases at the higher H$_{\rm CF}$ in case of the $x=$ 0.2 sample. As discussed above, the presence of larger FM cluster in the $x=$ 0.2 sample results in the lesser blocking of the spins at the FM--AFM interface, which lowers the value of M$_0$. Moreover, the high cooling field further enhance the FM interactions/grains causing the observed reduction in the M$_0$ for the $x =$ 0.2 sample for H$_{\rm CF} > $ 20~kOe.  

\begin{figure} 
\includegraphics[width=0.46\textwidth]{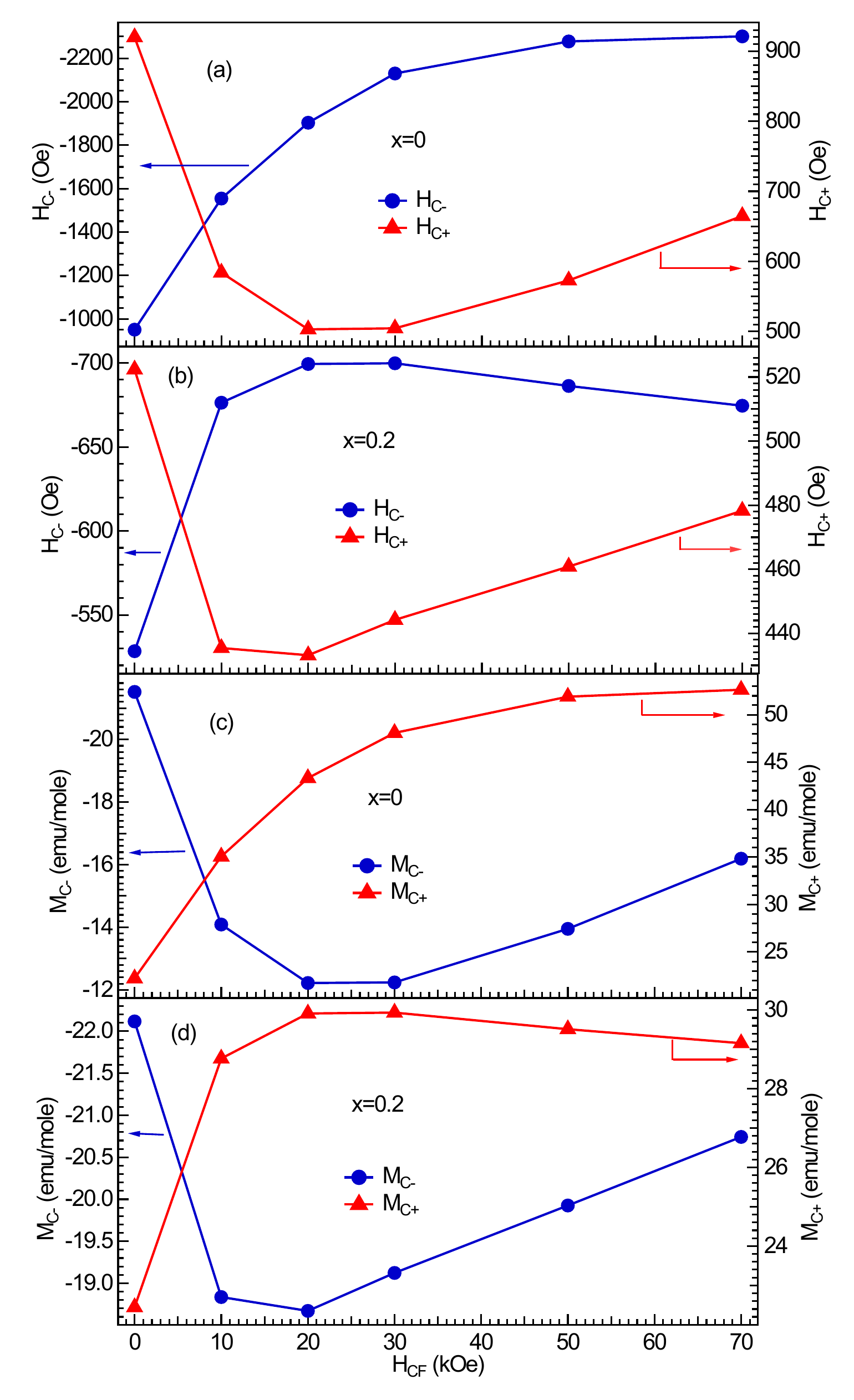}
\caption {(a, b) The cooling field dependence of the negative (on left axis) and positive (on right axis) coercive fields, and (c, d) the values of remanent magnetization, for both the samples.} 
\label{EB4_2}
\end{figure}

In order to further understand the effect of frozen spins on the EB phenomena, we plot the positive and negative coercivity, and remanent magnetization separately in Figs.~\ref{EB4_2}(a--d) for both the samples. It is interesting to note that in case of the $x=$ 0 sample, the negative (left) coercivity (H$_{\rm C-}$) value increases with increase in the cooling field up to 70~kOe, whereas the positive (right) coercivity (H$_{\rm C+}$) first decreases up to 20~kOe and then start increasing with further increase in the H$_{\rm CF}$, as shown in Fig. \ref{EB4_2}(a). This enhancement in the H$_{\rm C+}$ at the higher magnetic fields is responsible for the observed reduction in the H$_{\rm EB}$ for H$_{\rm CF}>$ 50~kOe in this sample [see Fig.~\ref{EB1}(a)]. On the other hand, the negative (positive) coercivity increases (decreases) up to around 20~kOe and then there is slight decrease (significant increase) in their values at H$_{\rm CF}>$ 20~kOe for the $x=$ 0.2 sample. This indicates that the rate of increment in the positive coercivity is higher than the reduction rate in the negative coercivity for H$_{\rm CF}>$ 20~kOe [see Fig.~\ref{EB4_2}(b)]. The larger FM domains in case of the $x=$ 0.2 as compared to the $x=$ 0 sample is responsible for the observed reduction in the M$_0$ and consequently decrease (increase) in the H$_{\rm C-}$ (H$_{\rm C+}$), and hence decay of the EB effect for the H$_{\rm CF}>$ 20~kOe. Moreover, the positive coercivity is found to be more responsible for such deterioration in the EB effect at the higher cooling fields for both the samples. Further, the M$_{\rm C-}$ and M$_{\rm C+}$ exhibit behavior analogous to that of H$_{\rm C+}$ and H$_{\rm C-}$, respectively [see Figs.~\ref{EB1}(c, d)]. In a typical field cycle of the M--H loop, the M$_{\rm C+}$ and M$_{\rm C-}$ are followed by the H$_{\rm C-}$ and H$_{\rm C+}$, respectively, resulting in their similar response to the blocked spins and, consequently, the H$_{\rm CF}$. 

\subsection{\noindent ~The training effect}

\begin{figure} 
\includegraphics[width=3.55in]{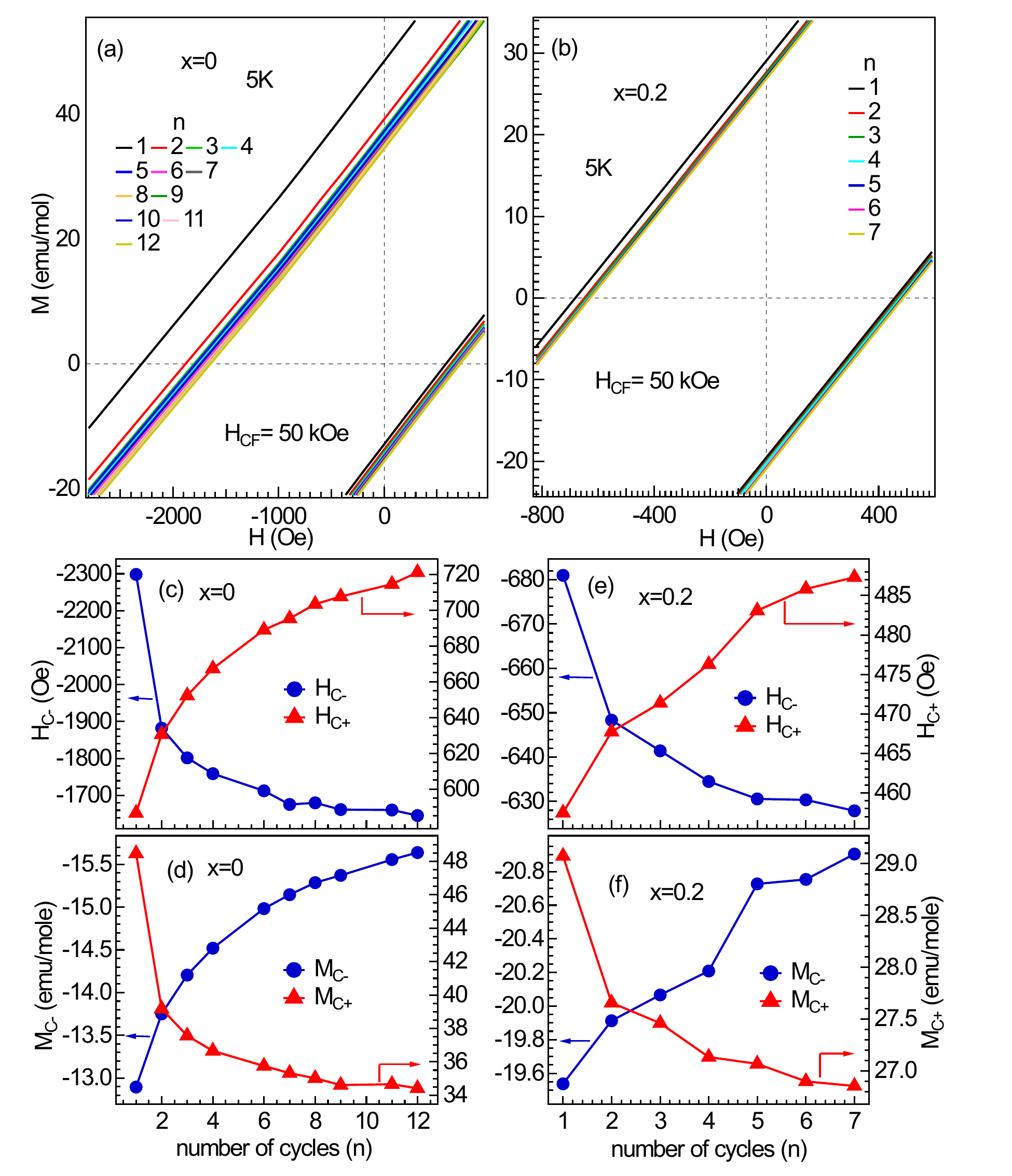}
\caption {The enlarged view of the M--H curves consecutively recorded after cooling the (a) $x=$ 0 and (b) 0.2 samples in 50~kOe magnetic field at 5~K. The variation in the negative (on left axis) and positive (on right axis) coercive fields and remanent magnetization with the number of field cycles for the (c, d) $x=$ 0 and (e, f) 0.2 samples, respectively.}
\label{Training}
\end{figure}

It is important to understand the evolution of the anisotropic exchange interactions and hence the EB parameters with the M-H cycles in these samples. Therefore, we perform the training effect measurements on both the samples, where we first cool the samples down to 5~K in the presence of 50~kOe magnetic field, and then record the consecutive M--H curves, as enlarged view is shown in Figs.~\ref{Training}(a) and (b) for the $x=$ 0 and 0.2 samples, respectively. A gradual reduction in the EB effect is clearly visible for both the samples, which can be understood in terms of the demagnetization of the AFM/CG spins at the magnetically disordered interface and hence decrease in the unidirectional anisotropy with the field cycles \cite{Keller_PRB_02}. The continuous cycling of the magnetic field results in the relaxation of the blocked spins at the domain boundaries. Note that the negative (positive) coercivity (remanent magnetization) changes much faster as compared to the positive (negative) branch for both the samples [see Figs.~\ref{Training}(a, b)]. However, this effect is more prominent in case of the $x=$ 0 sample as compared to the $x=$ 0.2. For the quantitative analysis, the positive and negative coercive field and remanent magnetization are plotted for both the samples in Figs.~\ref{Training}(c--f). These plots show that the negative (positive) coercivity (remanent magnetization) decays rapidly within the first 3-4 cycles and then show almost saturating behavior. On the other hand, positive (negative) coercivity (remanent magnetization) decays slowly and show a much lesser tendency to saturate even after several cycles. This behavior is consistent with the reports in Ref.~\cite{Karmakar_PRB_08, Shameem_JMMM_18, Wee_PRB_04}, where the asymmetric training effect is attributed to the dominance of the thermal activation in the FM region. The domains in the SG/AFM regions may get switched due to thermal activation along each branch of the loop \cite{Shameem_JMMM_18, Wee_PRB_04}.  

\begin{figure*} 
\includegraphics[width=7.1in]{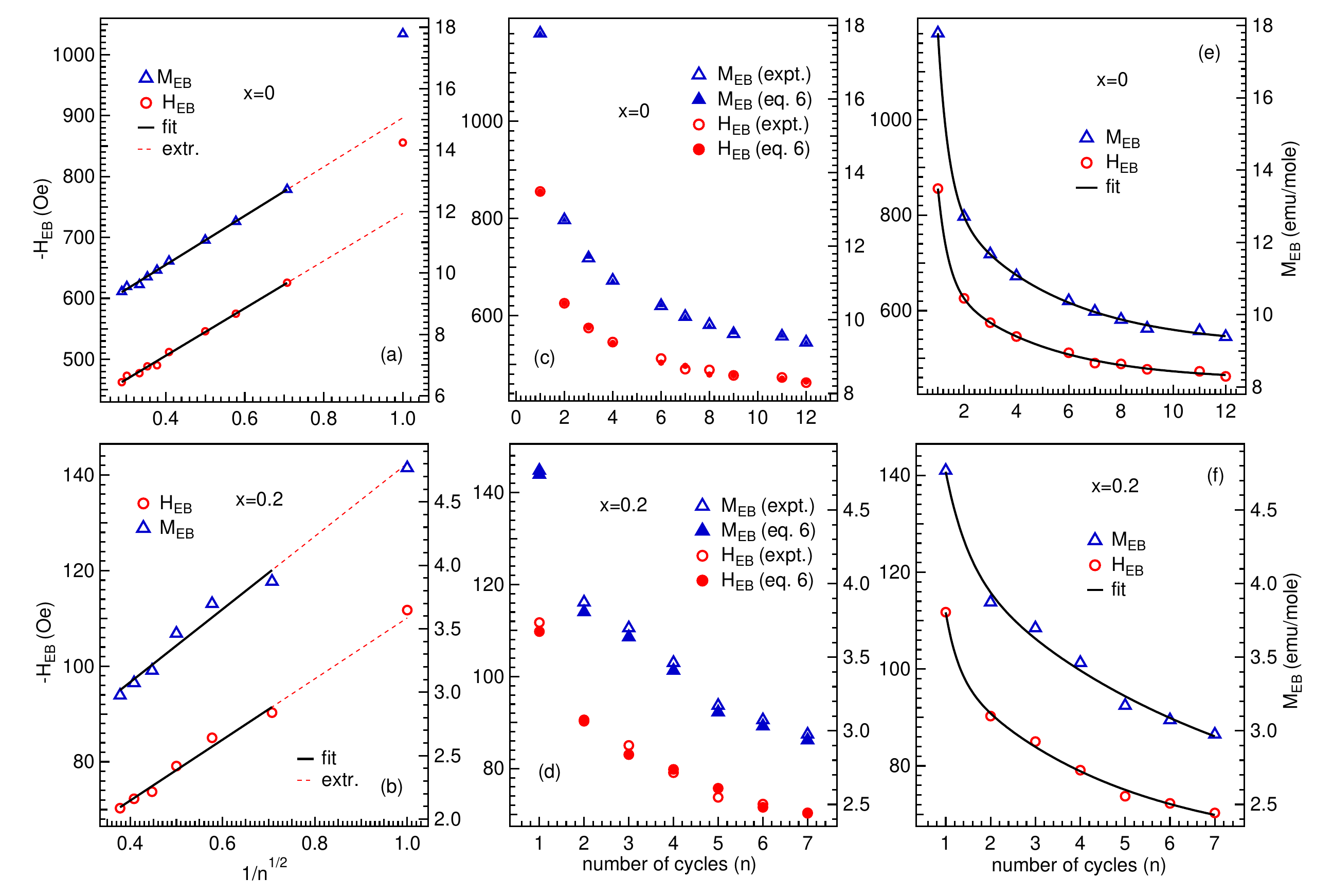}
\caption {(a, b) The EB field (H$_{\rm EB}$; on left axis) and EB magnetization (M$_{\rm EB}$; on right axis) versus 1/$\sqrt{n}$ plot for the $x=$ 0 and 0.2 samples, respectively. The black solid lines represent the power law fit of the data for $n>1$ and the red dashed lines represent the extrapolation of the data to $n=$ 1. (c, d) The dependence of H$_{\rm EB}$ and M$_{\rm EB}$ on the number of field cycles, where empty and solid symbols represent the experimental  and generated data from equations~\ref{BinekH} and \ref{BinekM}, respectively. (e, f) The best fit (solid black lines) of the H$_{\rm EB}$ and M$_{\rm EB}$ versus $n$ data using the equations~\ref{frozenH} and \ref{frozenM} for the $x=$ 0 and 0.2 samples, respectively.}
\label{Training2}
\end{figure*}

\begin{table*}
\label{tab_ac_fit}
\caption{The parameters extracted by fitting the training effect data of the $x=$ 0 and 0.2 samples using the equations~\ref{powerH}-\ref{frozenM}.}
\begin{tabular}{p{3cm}p{1.5cm}p{1.5cm}p{1.5cm}p{1.5cm}p{1.5cm} p{1.5cm}p{1.3cm}p{1.3cm}p{1.3cm}p{1.3cm}}
\hline
\hline
 fit using equation(s) & &$x=$ 0 & & &  & & & $x=$ 0.2\\
\hline
H$_{\rm EB}$/M$_{\rm EB}$ by \ref{powerH} \& \ref{powerM} & H$_{\rm EB}^\infty$ & k$_{\rm H}$ & M$_{\rm EB}^\infty$ & k$_{\rm M}$ & & & H$_{\rm EB}^\infty$ & k$_{\rm H}$ & M$_{\rm EB}^\infty$ & k$_{\rm M}$\\
 & -350(4) & 389(9) & 7.1(1) & 8.0(1) & & & -46(2) & 64(5) & 1.9(2) & 2.9(3)\\
 \hline
H$_{\rm EB}$/M$_{\rm EB}$ by \ref{BinekH} \& \ref{BinekM} & H$_{\rm EB}^\infty$ & $\gamma_{\rm H}$ & M$_{\rm EB}^\infty$ & $\gamma_{M}$ & & & H$_{\rm EB}^\infty$ & $\gamma_{\rm H}$ & M$_{\rm EB}^\infty$ & $\gamma_{M}$\\
 & -281(18) & 1.21x10$^{-6}$ & 5.6(1) & 2.8x10$^{-3}$ & & & -40(6) & 5.8x10$^{-5}$ &  1.7(4) & 3.0x10$^{-2}$\\
\hline
H$_{\rm EB}$ by \ref{frozenH} & H$_{\rm EB}^\infty$ & A$_f$ & P$_f$ & A$_r$ & P$_r$ & H$_{\rm EB}^\infty$ & A$_f$ & P$_f$ & A$_r$ & P$_r$\\
&-454(7) & -2498(47) & 0.39(9) & -264(28) &  3.8(6) & -62(5) & -420(11) & 0.28(2) & -48(3) & 3.7(7) \\
M$_{\rm EB}$ by \ref{frozenM} & M$_{\rm EB}^\infty$ & A$_f$ & P$_f$ & A$_r$ & P$_r$ & M$_{\rm EB}^\infty$ & A$_f$ & P$_f$ & A$_r$ & P$_r$\\
&9.1(1) & 54(16) & 0.41(6) &5.2(3) & 4.1(5) & 2.3(2) & 5.5(1) & 0.5(1) & 2.2(2) & 6.1 (1)\\
\hline
\hline
\end{tabular}
\end{table*}

To quantitatively understand the nature of these glassy magnetic interactions, first we fit the variation of the EB field and the EB magnetization with the number of recorded M--H loops using the power law \cite{Paccard_PSS_66}, as below: 
\begin{equation}
H_{EB}^n - H_{EB}^\infty= \frac{k_{\rm H}}{\sqrt{n}} \;\;\;\;\; {\rm for}\; n>1
\label{powerH}
\end{equation}
\begin{equation}
M_{EB}^n - M_{EB}^\infty= \frac{k_{\rm M}}{\sqrt{n}} \;\;\;\;\; {\rm for}\; n>1, 
\label{powerM}
\end{equation}
where $n$ is the number of recorded M--H loops, H$_{EB}^\infty$  and M$_{EB}^\infty$ are the values of H$_{\rm EB}$ and M$_{\rm EB}$ for $n=\infty$, and k$_{\rm H}$ and k$_{\rm M}$ are the constants. Figs.~\ref{Training2}(a, b) show the variation in H$_{\rm EB}$ (left axis) and M$_{\rm EB}$ (right axis) with $1/\sqrt{n}$ for the $x=$ 0 and 0.2 samples, respectively, where the black solid lines represent the best fit using the power law for the $n>1$, as equations~\ref{powerH} and \ref{powerM} are not valid for the first cycle. The best fit parameters are summarized in Table I, which show a significant decrease in the H$_{EB}^\infty$ (M$_{\rm EB}^\infty$) and $k_{\rm H}$ ($k_{\rm M}$) with the La substitution, indicating a reduction in the EB effect. Further, the power law underestimate the values of both H$_{\rm EB}$ and M$_{\rm EB}$ for $n=$ 1, as shown by the red dashed lines in Figs.~\ref{Training2}(a, b) for the $x=$ 0 and 0.2 samples, respectively. However, this effect is more prominent in case of the $x =$ 0 sample. A steep decrease in the H$_{\rm EB}$ or M$_{\rm EB}$ between the first and second cycles is usually related with the relaxation of the spins at the disordered interfaces due to the surface drag from the switching field, whereas a gradual decrease in the exchange bias parameters is attributed to the relaxations of the thermally activated spins during the consecutive field cycles and follow the power law behavior \cite{Shameem_JMMM_18, Outon_JMMM_06}. A smaller deviation in the EB parameters from the power law at $n=$ 1 in case of the $x=$ 0.2 sample indicate the more thermal relaxation of the pinned spins during the initial field cycles as compared to the $x=$ 0 sample. 

\begin{figure} 
\includegraphics[width=3.6in]{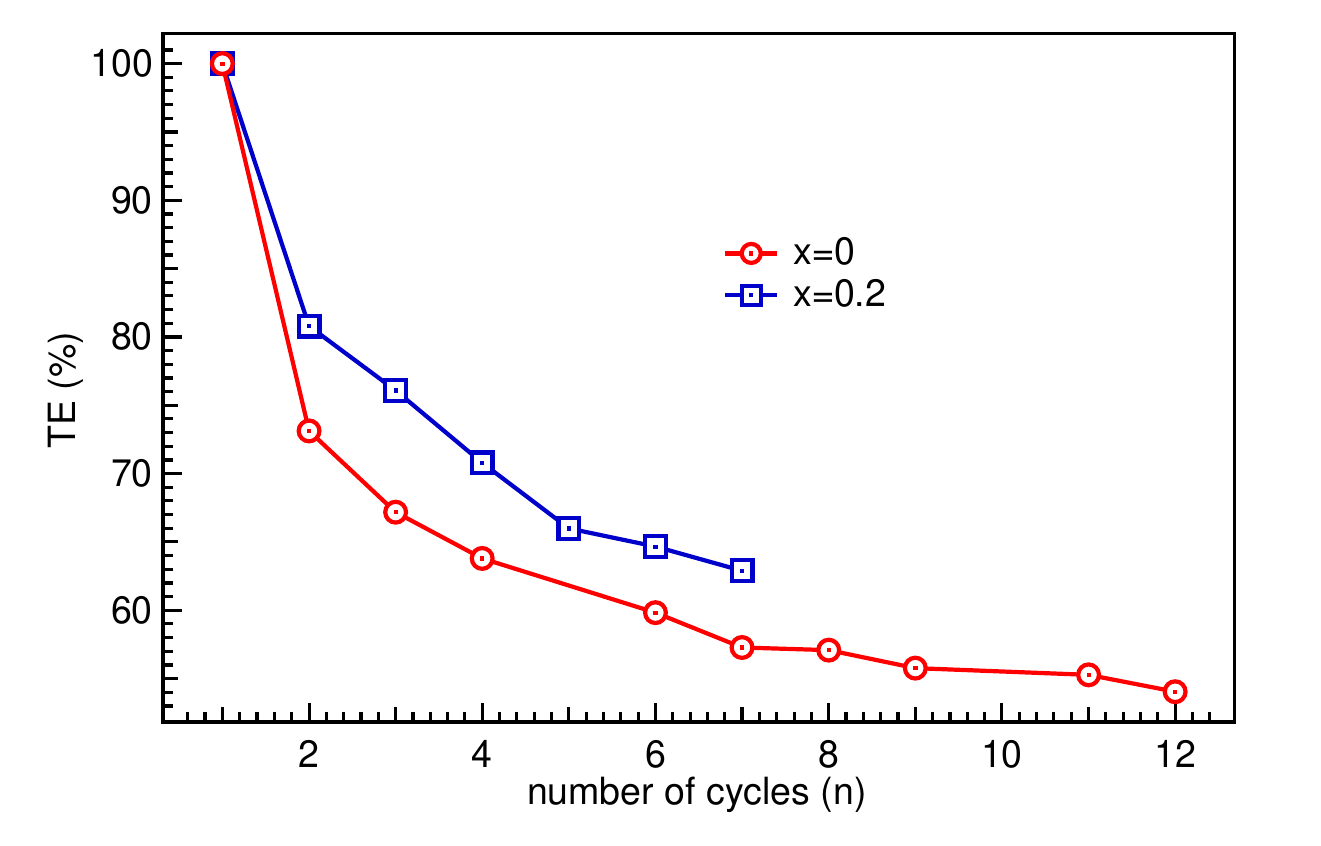}
\caption {The relative percentage reduction in the training effect (equation~\ref{TEdecay}) with the number of field cycles ($n$) for the $x=$ 0 and 0.2 samples.} 
\label{Training3}
\end{figure}

\begin{figure*} 
\includegraphics[width=7.2in]{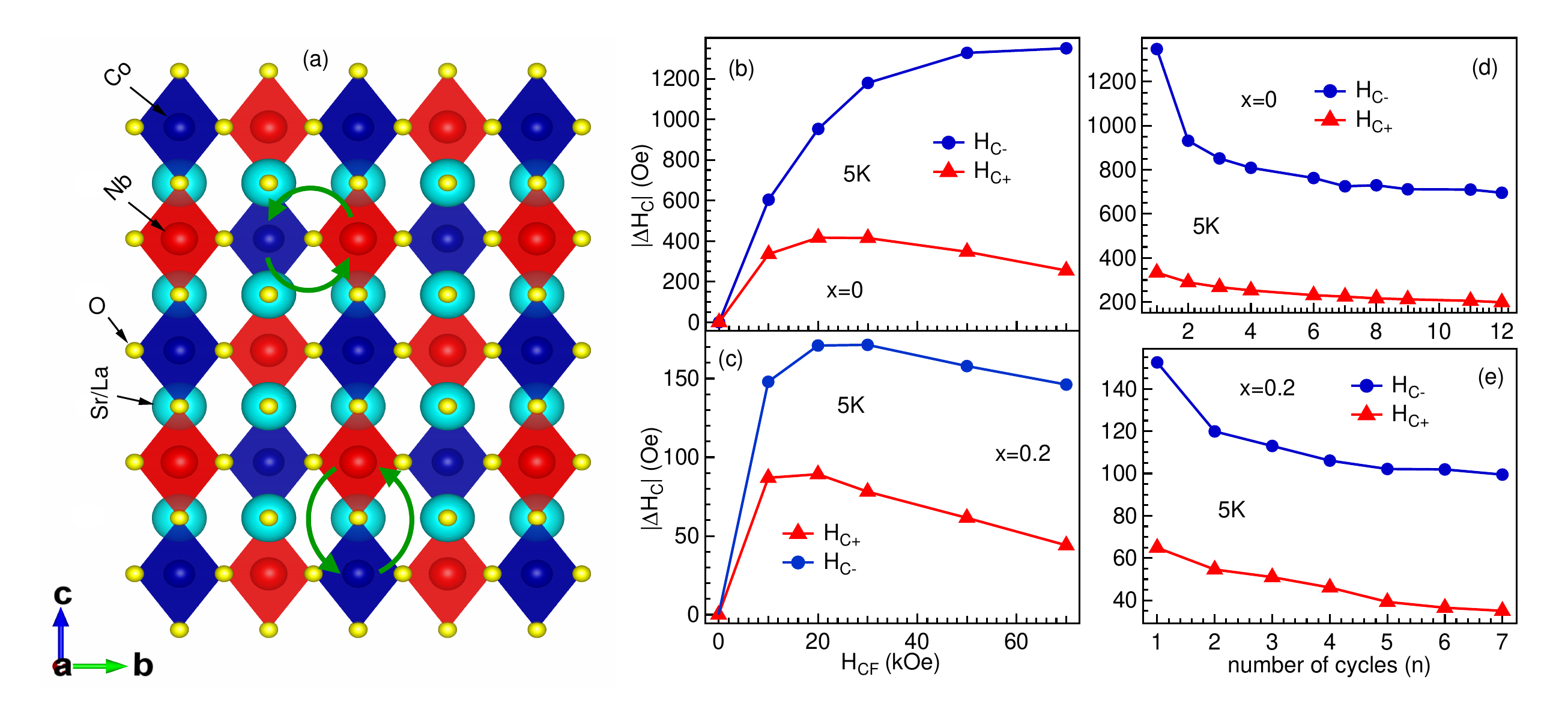}
\caption {(a) A 3D schematic of the Sr$_2$CoNbO$_6$ crystal structure where arrows indicate the possible disorder paths in the crystal. (b, c) The variation of the position and negative coercivity with respect to the zero field coercivity as function of cooling field, and (d, e) the number of field cycles after cooling the samples in 50~kOe field at 5~K for the $x=$ 0 and 0.2 samples, respectively.}
\label{Final}
\end{figure*}

The training effect data can be analyzed using the spin relaxation model proposed by C. Binek \cite{Binek_PRB_04}, which suggest the demagnetization of the non-FM spins at the disordered interface with the field cycling. In this model, the H$_{\rm EB}$ and M$_{\rm EB}$ can be expressed as a function of $n$ using the following recursive formulae \cite{Binek_PRB_04} 
\begin{equation}
H_{\rm EB} (n+1) - H_{\rm EB} (n)=-\gamma_H [H_{\rm EB} (n)-H_{EB}^\infty]^3
\label{BinekH}
\end{equation}
\begin{equation}
M_{\rm EB} (n+1) - M_{\rm EB} (n)=-\gamma_M [M_{\rm EB} (n)-M_{EB}^\infty]^3,
\label{BinekM}
\end{equation}
where  $\gamma_H$ and $\gamma_M$ are the constants. The H$_{\rm EB}$ and M$_{\rm EB}$ curves are generated using the above equations by taking the experimental H$_{\rm EB}$(1) and M$_{\rm EB}$(1) data, and the values of $\gamma_H$, $H_{EB}^\infty$, $\gamma_M$, and $M_{EB}^\infty$ as given in Table I. These results are presented by the solid symbols in Figs.~\ref{Training2}(c) and (d) for the $x=$ 0 and 0.2 samples, respectively, where a good agreement between the experimental and the generated data [using equations~\ref{BinekH} and \ref{BinekM}] is clearly observed even for the $n=$ 1. Also, we observe a reduction in the values of H$_{EB}^\infty$ and M$_{\rm EB}^\infty$ with the La substitution analogous to the power law, discussed above. Further, the values of $\gamma_{\rm H}$ and $\gamma_{\rm M}$ significantly increase with the La substitution. Now, by putting $\gamma_{\rm H}$ = b/K$^2\xi$, where $b$ and $\xi$ are different constants, and K = --JS$_{\rm FM}$/M$_{\rm FM}$t$_{\rm FM}$ in equation~\ref{EB}, we can write H$_{\rm EB}$ =KS$_{\rm AFM}$/$\mu_0$ \cite{Binek_PRB_04, Meiklejohn_PR_56, Karmakar_PRB_08}. Thus, the observed enhancement in the value of $\gamma_{\rm H}$ indicate the lower value of K, and hence the H$_{\rm EB}$ value in case of the $x=$ 0.2 sample. However, this model slightly underestimate the values of H$_{EB}^\infty$ and M$_{\rm EB}^\infty$ as compared to that extracted from the power law (see Table~I), which is also observed in Ref.~\cite{Shameem_JMMM_18}. 

Therefore, we further analyze the training effect data using the model proposed in Ref.~\cite{Mishra_PRL_09}, which consider the two different types of uncompensated spins, namely frozen and rotating spins at the disordered interface having different relaxation times, where H$_{\rm EB}$ and M$_{\rm EB}$ can be expressed as a function of $n$ as below: 
\begin{equation}
H_{\rm EB}^n=H_{\rm EB}^\infty+A_f exp\left(\frac{-n}{p_f}\right)+A_r exp\left(\frac{-n}{p_r}\right),
\label{frozenH}
\end{equation}
\begin{equation}
M_{\rm EB}^n=M_{\rm EB}^\infty+A_f exp\left(\frac{-n}{p_f}\right)+A_r exp\left(\frac{-n}{p_r}\right),
\label{frozenM}
\end{equation}
here the pre exponent term $A$ represents the weight factors having dimensions of magnetic field, the dimensionless parameters $p$ are the measure of the relaxation times, and the subscripts $f$ and $r$ represent the frozen and rotatable spin components, respectively. The solid black lines in Figs.~\ref{Training2}(e, f) show the best fit of H$_{\rm EB}$ and M$_{\rm EB}$ curves for the $x=$ 0 and 0.2 samples, respectively, using the above model and the fitting parameters are given in Table~I. We find that the weight factors A$_f$ and A$_r$ decreases with the La substitution, which indicate the reduction in the concentration of both frozen as well as rotatable spins at the disordered interface with $x$. Further, it can be inferred from Table~I that the rotatable components relax 10 and 13 times faster than the frozen spin component for the $x=$ 0 and 0.2 samples, respectively. This indicates that the rotatable spins play a dominating role in governing the EB effect as a function of $n$. Now, we calculate the relative percentage of reduction in the EB effect [TE(\%)] with the number of field cycles ($n$) using the following relation \cite{Ventura_PRB_08, Anusree_PRB_20}:
\begin{equation}
{\rm TE(\%)}=\left[1-\frac{(H_{\rm EB}^1-H_{\rm EB)}^n}{H_{\rm EB}^1}\right] \times 100, 
\label{TEdecay}
\end{equation}
In Fig.~\ref{Training3} we find that the TE(\%) value decreases to the 73\% and 81\% of its initial value only after the first field cycling ($H_{\rm EB}^2$) in case of the $x=$ 0 and 0.2 samples, respectively. It is clear from Fig.~\ref{Training3} that despite the significantly lower $H_{\rm EB}$ [see Fig.~\ref{EB1}(c)], the $x=$ 0.2 sample shows the much stable EB effect with the field cycling as compared to the $x=$ 0 sample. For example, the H$_{\rm EB}$ reduces to the 57\% of its initial value ($H_{\rm EB}^1$) in case of the $x=$ 0 sample, whereas retains 63\% in the $x=$ 0.2 sample at the 7$^{\rm th}$ cycle. The lower effective FM--AFM disordered interface decrease the probability of the relaxation of uncompensated spins with the field cycles, resulting in the relatively robust EB effect in case of the $x=$ 0.2 as compared to the $x=$ 0 sample. 

\subsection{\noindent ~Origin and evolution of the EB effect}

It is important to mention that the Co is present predominantly in 3+ valence state in the $x =$ 0 sample \cite{Kumar_PRB3_22}, which excludes the possibility of the competing magnetic interactions due to the multiple valence states of Co as the origin of the observed EB effect in this sample. The XRD, EXAFS, and ND measurements indicate that Co$^{3+}$ and Nb$^{5+}$ (non-magnetic) ions can swap between their given Wyckoff positions, as shown in  Fig.~\ref{Final}(a), resulting in the Co$^{3+}$-O-Co$^{3+}$ exchange interactions in case of the $x=$ 0 sample \cite{Kumar_PRB1_20, Azcondo_DT_15, Kumar_PRB3_22}. Moreover, the high-resolution electron microscopic measurements reported in Ref.~\cite{Azcondo_DT_15} suggest the presence of some ordered domains at nanoscale range, which give rise to the additional Co$^{3+}$-O-Nb$^{5+}$-O-Co$^{3+}$ channels. Further, depending on the relative strength of the crystal field and Hund's exchange energy, the Co$^{3+}$ can exist in the different spin states, namely, low spin (LS, t$_{2g}^6$e$_g^0$; S=0), intermediate spin (IS, t$_{2g}^5$e$_g^1$; S=1) and/or high spin (HS, t$_{2g}^4$e$_g^2$; S=2) state, resulting in the complex magnetic interactions in this sample \cite{Kumar_PRB1_20}. For example, Goodenough–Kanamori rules predict LS Co$^{3+}$--O--HS Co$^{3+}$ interactions as FM and HS Co$^{3+}$--O--HS Co$^{3+}$ as AFM in nature \cite{Goodenough_PR_55, Kanamori_JPCS_59}. Therefore, the presence of these competing magnetic interactions of varying sign and strength arising from the different spin-states of Co$^{3+}$ give rise to the frustration in the spins and consequently cluster glass like behavior and EB effect in the $x = $ 0 sample having only one magnetic atom with a fix valence state \cite{Kumar_PRB2_20}. 

However, the substitution of La$^{3+}$ ions at the Sr$^{2+}$ site will transform the Co from 3+ to 2+ state in the same proportion \cite{Kumar_PRB3_22} and monotonically enhance the B-site ordering, and hence the Co$^{3+/2+}$-O-Nb$^{5+}$-O-Co$^{3+/2+}$ interaction channels in Sr$_{2-x}$La$_x$CoNbO$_6$ samples \cite{Kumar_PRB1_20, Kumar_PRB2_20, Kumar_PRB3_22, Kumar_JPCL_22}. We observe the reduction in the coercivity, but at the same time an enhancement in the saturation magnetization (M$_{\rm S}$) values in the $x=$ 0.2 as compared to the $x=$ 0 sample, see the insets of Figs.~\ref{EB1}(a, b). An increase in the size of the FM cluster, as discussed above, may be responsible for the observed enhancement in the M$_{\rm S}$ value, whereas a decrease in the coercivity suggest the reduction in the strength of these FM interactions in the $x=$ 0.2 sample. Recently, a significantly larger EB effect is reported in SrLaFe$_{0.5}$(Mn$_{0.25}$Co$_{0.25}$)O$_4$ as compared to the SrLaFe$_{0.25}$(Mn$_{0.25}$Co$_{0.5}$)O$_4$ in spite of the smaller coercivity of the former, indicating the dominating contribution of the effective CG-AFM interface in governing the EB effect \cite{Anusree_PRB_20}. The larger FM cluster reduces the frustrated spins at the CG-FM/AFM interfaces and hence lower EB effect in the $x=$ 0.2 sample. Moreover, we recently report that the size of the glassy spin cluster decreases with the La substitution in Sr$_{2-x}$La$_x$CoNbO$_6$ samples for the $x \leqslant$0.4 \cite{Kumar_PRB2_20}, which can also play a key role in reducing the EB effect in case of the $x=$ 0.2. The larger FM and smaller CG domains also reduce the critical cooling field (H$_{\rm CF}^{\rm crit}$), above which the H$_{\rm EB}$ starts deceasing, from 50~kOe in case of the $x=$ 0 to 20~kOe for the $x=$ 0.2 sample. 

More importantly, we observe the asymmetric exchange bias effect in both the field cooled M--H curves with varying H$_{\rm CF}$ as well as in the training effect measurements. The reduction in the EB parameters at the higher cooling field was found to be mainly governed by the H$_{\rm C+}$ (M$_{\rm C-}$) branch; on the other hand, the training effect was controlled by the H$_{\rm C-}$ (M$_{\rm C+}$) branch. In order to understand this, we plot the magnitude of change in the positive and negative zero field coercivity with respect their field cooled values (shift in the individual branches) as a function of H$_{\rm CF}$ and $n$ for both the samples in Figs.~\ref{Final}(b--e). It is interesting to note that the shift in the negative branch in the field cooled M-H is significantly higher as compared to the positive branch in both the samples [see Fig. \ref{Final}(b, c)]. The lower shifting of the positive branch is attributed to the relaxation of the locked spins during the first half of the field cycling. Moreover, a significantly higher shift in the negative coercivity signifies that the enhancement in the H$_{\rm C}$ due to enhancement (growth) in the ferromagnetic interactions (domains) with the cooling field is mainly attributed to the H$_{\rm C-}$. In case of the $x=$ 0 sample, the change in negative coercivity increases with increase in the cooling field up to 70~kOe, whereas the positive coercivity decreases for H$_{\rm CF}>$20~kOe [see Fig.~\ref{Final}(b)]. This clearly indicates the different influence of the cooling field on the two branches. On the other hand, the change in both the coercivities decreases for H$_{\rm CF}>$20~kOe in case of the $x=$ 0.2 sample [see Fig.~\ref{Final}(c)], i.e, the M--H loops start shifting back to the right side for H$_{\rm CF}>$20~kOe in this sample. This effect is more prominent in the positive branch. Further, we observe an unequal shift of the two branches of the M--H loop and their different evolution with the field cycles, as shown in Figs.~\ref{Final}(d, e), which is responsible for the asymmetric training effect observed in both the samples. However, this effect is less prominent in case of the $x=$ 0.2 as compared to the $x=$ 0, which results in the more stable EB effect in the former. 

\section{\noindent ~Conclusions}

The influence of antisite disorders on the EB effect of Sr$_{2-x}$La$_x$CoNbO$_6$ ($x=$ 0, 0.2) samples is investigated in detail using the FC M--H and training effect data analysis. The field dependence of the T$_{\rm irr}$ deviates from both the AT and GT lines, which suggest the presence of different universality class with the moderate unidirectional exchange interactions in the $x=$ 0 sample. A significant reduction in the EB effect is observed in case of the $x=$ 0.2 sample due to the larger FM and smaller CG domains as compared to the $x=$ 0 sample. The EB effect show a notable reduction at the higher H$_{\rm CF}$ for both the samples and the positive (negative) coercivity (remanent magnetization) is found to predominantly govern this effect. Further, both the samples show the asymmetric training effect, which is mainly governed by the negative coercivity branch. The analysis of the decay in the EB effect with the M--H cycles shows that the rotatable spins relax 10 and 13 times faster than the frozen spins at the disordered interfaces of the $x=$ 0 and 0.2 samples, respectively. Through the training effect measurements, we reveal a much more stable EB effect with the M--H cycles for the $x=$ 0.2 sample as compared to that of the $x=$ 0. The detailed investigation of the low temperature spin dynamics responsible for systematic tunning of the EB effect with the La substitution in these samples can be useful to engineer the suitable candidates for the desired applications. 

\section{\noindent ~Acknowledgments}

A.K. acknowledges the UGC for the fellowship and the Department of Physics of IIT Delhi for providing PPMS facility to perform the detailed magnetization measurements. R.S.D. is grateful to the SERB--DST for the financial support through a core research grant (project reference no. CRG/2020/003436).

\end{document}